# Enhancing LSPR sensitivity of Au gratings through graphene coupling to Au film

T. Maurer,[1,*] R. Nicolas,[1,4] G. Lévêque,[2] P. Subramanian,[3] J. Proust,[1] J. Béal,[1] S. Schuermans,[1] J.-P. Vilcot,[2] Z. Herro,[4] M. Kazan,[5] J. Plain,[1] R. Boukherroub,[3] A. Akjouj,[2] B. Djafari-Rouhani,[2] P.-M. Adam,[1] and S. Szunerits[3,*]

1. Laboratoire de Nanotechnologie et d'Instrumentation Optique, ICD CNRS UMR STMR 6279, Université de Technologie de Troyes, CS 42060, 10004 Troyes, France

2. Institut d'Electronique, de Microélectronique et de Nanotechnologie (IEMN, CNRS-8520), Cité Scientifique, Avenue Poincaré, 59652 Villeneuve d'Ascq, France

3. Institut de Recherche Interdisciplinaire (IRI), USR-3078, Université Lille 1, 50 Avenue de Halley, BP 70478, Villeneuve d'Ascq 59658, France.

4. Université Libanaise, EDST, Platforme de Recherche en NanoSciences et NanoTechnologie PR2N, Fanar, BP 90239, Lebanon

5. Department of Physics, American University of Beirut, Riad El-Solh 1107 2020, Beirut, Lebanon

A particular interesting plasmonic system is that of metallic nanostructures interacting with metal films. As the LSPR behavior of gold nanostructures (Au NPs) on the top of a gold thin film is exquisitely sensitive to the spacer distance of the film-Au NPs, we investigate in the present work the influence of a few-layered graphene spacer on the LSPR behavior of the NPs. The idea is to evidence the role of few-layered graphene as one of the thinnest possible spacer. We first show that the coupling to the Au film induces a strong lowering at around 507nm and sharpening of the main LSPR of the Au NPs. Moreover, a blue shift in the main LSP resonance of about 13 nm is observed in the presence of a few-layered graphene spacer when compared to the case where gold nanostructures are directly linked to a gold thin film. Numerical simulations suggest that this LSP mode is dipolar and that the hot spots of the electric field are pushed to the top corners of the NPs, which makes it very sensitive to surrounding medium optical index changes and thus appealing for sensing applications. A figure of merit (FoM) of such a system (gold/graphene/ Au NPs) is 2.8, as compared to 2.1 for gold/Au NPs either a 33% sensitivity gain and opens up new sensing strategies.

The phenomenon of localized surface plasmon resonance (LSPR) has been extensively studied over the last decade (Mayer and H. Hafner, 2011, Szunerits and Boukherroub, 2012). Because of intense local electrical field enhancements and sharp resonance excitation peaks, metallic nanoparticles are of great interest for the development of chemical and biological sensors as well as their use as signal enhancers in surface-based spectroscopies (Haes and Van Duyne, 2002, Xu et al., 2012b). A particular interesting plasmonic system that has received somewhat less attention is that of metallic nanostructures interacting with metal films (He et al., 2004, Tokareva et al., 2004, Cesario et al., 2005, Leveque and Martin, 2006, Levêque and Martin, 2006, Mock et al., 2008, Chu and Crozier, 2009, Hohenau and Krenn, 2010, Mock et al., 2012). This system has been predicted to display a wealth of interesting optical phenomena due to the complex interaction of the confined LSPR properties of the particles with the delocalized thin film surface plasmon polariton. Numerical (Leveque and Martin, 2006, Levêque and Martin, 2006) as well as experimental results have been presented by several groups showing the distance-dependent plasmon resonant coupling (Mock et al., 2008, Mock et al., 2012). Mock and collaborators investigated the distance-dependent coupling between spherical Au NPs (60 nm in diameter) and a gold film (45 nm in thickness) by using polyelectrolyte assemblies with varying thickness (0-22.5 nm) as the spacer between the Au film and the Au NPs (Mock et al., 2008). By characterizing the scattering of a single nanoparticle, it was shown that when the nanoparticle is in close proximity to the metal surface, damping of the horizontal (parallel to the Au film surface) particle LSPR mode results in vertically (perpendicular to the Au film surface) polarized NP scattering and a doughnut-shaped far field image. Cesario et al. showed with transmission measurements performed on a Au film coated with an indium tin oxide (ITO) spacer layer of 20 nm, onto which an ordered array of Au NPs (20 nm in diameter) was deposited, that two plasmonic modes are apparent: a band at lower wavelength (around 700nm) attributed to the LSPR of the isolated NPs and a second plasmonic band at higher wavelength (above 800nm) resulting from the excitation of the surface plasmon polariton (SPP) branch (1,0) by grating coupling (Cesario et al., 2005). These results were confirmed by reflection light extinction measurements on a similar system using a $SiO_2$ spacer, which in addition put into evidence a LSPR mode at shorter wavelength, independent of the NPs' diameter and attributed to the excitation of the (1,1) SPP mode of the Au film (Chu and Crozier, 2009). Recently, Krenn and co-workers revealed a period independent extinction band at 520 nm in addition to a band at 600 nm, which shifts to larger wavelengths for larger array periods on an interface consisting of rectangular Au NP grating directly deposited onto a 25 nm thick gold film (Hohenau and Krenn, 2010). Numerical simulations indicated that for such small array periods with interparticle distance inferior to 500 nm, symmetric SPP modes cannot be excited. The LSPR mode at 520 nm mode was thus attributed to a combination of vertically oriented dipole LSPR located at the NPs and scattering to high-energy SPP.

---

[*] To whom correspondence should be send to: Thomas Maurer (thomas.maurer@utt.fr); Sabine Szunerits (sabine.szunerits@iri.univ-lille1.fr)

Motivated by previous work showing that the LSPR behavior of metallic nanostructures on the top of a metal thin film is exquisitely sensitive to the spacer distance of the film-NPs (Mock et al., 2008), we investigate in the present work the influence of a few-layered graphene spacer. The interest of graphene for plasmonic devices has been highlighted in several recent papers (Salihoglu et al., 2011, Reed et al., 2012, Xu et al., 2012a, Szunerits et al., 2013). Graphene has been considered as an alternative coating for silver (Choi et al., 2011, Szunerits et al., 2013) and gold (Wu et al., 2010) based SPR as it is believed to have several advantages: (i) graphene has a very high surface to volume ratio, which is expected to be beneficial for efficient adsorption of biomolecules as compared to naked gold; (ii) graphene is expected to increase the adsorption of organic and biological molecules as their carbon-based ring structure allows π-stacking interaction with the hexagonal cells of graphene; (iii) controlling the number of graphene layers transferred onto the metal interface should allow tuning the SPR response and the sensitivity of SPR measurements (Wu et al., 2010). However, the LSPR behaviour of metallic nanostructures on the top of a metal thin film with a graphene spacer in-between has not been investigated so far. In this work, we take advantage of the two-dimensional structure of graphene with a thickness of 0.34 nm (Gupta et al., 2006) as a high optical index non dielectric spacer between a flat Au film and a Au NP array. The plasmonic properties of this interface are compared to Au NPs directly onto a 50 nm thick Au film. Moreover, the potential of such substrates for sensing applications is assessed and trilayer graphene is evidenced to enhance sensitivities of LSP sensors

The newly designed interface consists of Au NP gratings, produced by e-beam lithography (EBL), deposited onto graphene coated 50 nm Au film (**Fig. 1A**). The plasmonic properties of this interface are compared to similar Au NPs directly deposited onto a 50 nm Au film (**Fig. 1B**). Structural characterizations as well as SPR measurements evidenced good quality of transferred graphene and a thickness of 1.02nm which corresponds to three monolayers (see Supporting Information, Figure S1 and S2).

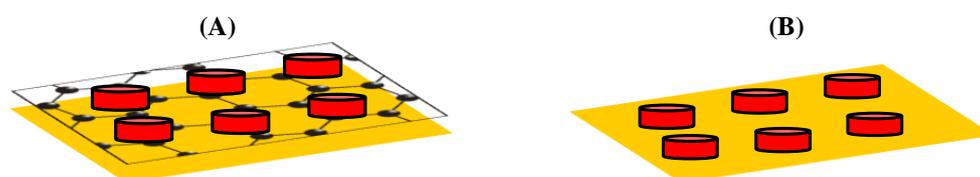

**Fig. 1** Schematic illustration of the different interfaces investigated: (A) graphene coated Au thin film decorated with Au NPs array; (B) Au NPs array directly deposited onto thin Au film without the graphene spacer layer.

The graphene-modified SPR interface was in the following decorated with an ordered array of Au NPs *via* electron beam lithography (EBL) (McCord and Rooks, 1997). On top of a chromium adhesion layer (d=3 nm), Au NPs of 50 nm in height, with a centre-to-centre distance of 300 nm and varying particle diameter (80 nm, 110 nm and 140 nm) were formed by EBL. The lithographically fabricated particles have roughly a cylindrical shape as seen from the scanning electron microscopy (SEM) images in **Fig. 2**.

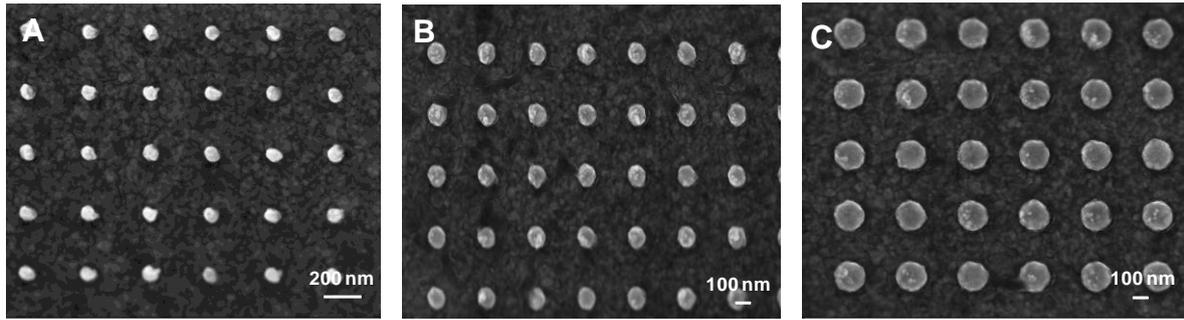

**Fig. 2.** SEM images of graphene-based SPR decorated with Au NPs by EBL with center-to-center distance of 300 nm. The particles are 50 nm in height and 80 nm (A), 110 nm (B), 140 nm (C) in diameter.

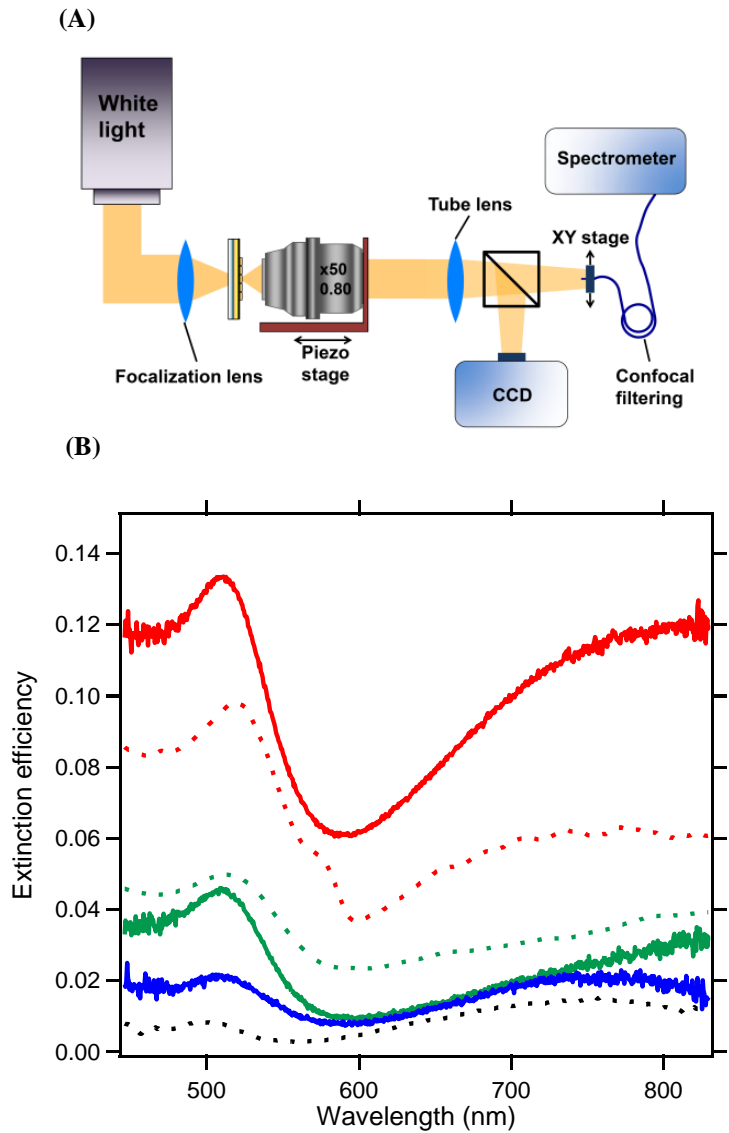

**Fig. 3**: (A) Set up for optical measurements. (B) Extinction spectra measured in air of the Au surface (dashed lines) and graphene-modified Au surface (full lines) decorated with Au NPs of 50 nm (black), 80 nm (blue), 110 nm (green) and 140 nm (red) in diameter, 50 nm in height and center-to-center distance of 300 nm. The signal was collected with a ×10 objective with a numerical aperture of NA = 0.15. The reference for calculating the extinction is taking on the gold film outside the arrays. The optical extinction spectrum of 80 nm Au NPs directly fabricated on Au film could not be resolved.

The extinction spectra of the systems have been measured with a transmission optical microscope coupled to a micro-spectrometer using a multimode optical fibre as confocal filtering, as schematically described in **Fig. 3A**. A ×10 objective lens (NA=0.15) allows for a detection area of ≈50 µm². **Fig. 3B** displays the extinction spectra under normal incidence for the different gratings fabricated onto graphene coated Au surface (full lines). Each of the curves is characterized by a sharp resonance peak at $\lambda_1$= 507 nm and a second band at higher wavelength, $\lambda_2$= 770 nm, which is however rather broad and not well defined for all the investigated interfaces. The position of the resonance band at shorter wavelength, $\lambda_1$, is size-independent due to the gold interband transitions and its full width at half maximum (fwhm) is decreasing with increasing the diameter of the NPs to reach fwhm≈50 nm for NPs of 140 nm in diameter. For comparison, the peak fwhm for the band at $\lambda_2$ is about 250 nm for this array. The position of the $\lambda_1$=507nm peak is a low wavelength mode compared to the one expected for Au NP gratings on glass substrates.

To understand the origin of the plasmonic band blue shift, a similar system without any spacing layer (graphene) between the Au NP gratings and the Au film was constructed (**Fig. 1B**) and the experimentally obtained extinction spectra are displayed in **Fig. 3B**. In the case of the $\lambda_1$ plasmon band the peak position is not size independent any longer, since the LSP resonance is at 505 nm for 50 nm NPs, 514 nm for 110 nm NPs and 520 nm for 140 nm NPs. The optical extinction spectrum of 80 nm Au NPs directly fabricated on Au film could not be resolved. However, the low wavelength LSP mode seems to be quite similar in both cases. Note that the graphene layers leads to little sharpening of the $\lambda_1$ plasmon band since its width is increased to 64 nm for Au NPs deposited directly on the Au film without any graphene spacer.

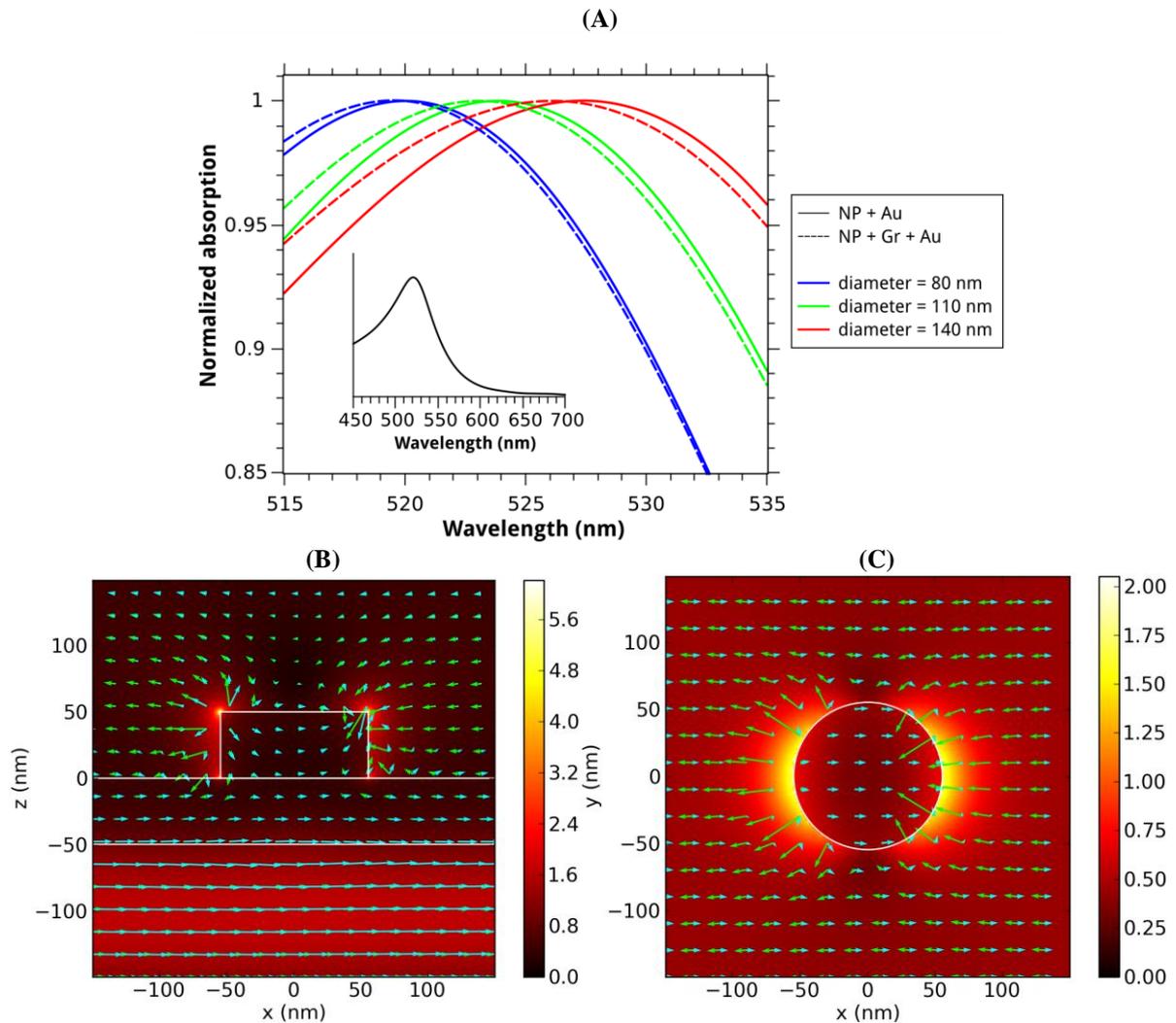

**Fig. 4**: (A) Computed extinction spectra of a single cylinder particle, diameter 80 nm (blue), 110 nm (green) and 140 nm (red), thickness 50 nm, placed on the Au surface (solid lines), or on the graphene-modified Au surface (dashed lines). (B) Computed distribution of the electric field inside a vertical section of the 110-nm-diameter particle on the Au substrate, at the resonant wavelength λ = 524 nm. Color scale: electric-field time-averaged amplitude, normalized to the incident plane wave amplitude; green vectors: electric-field real part; cyan vectors:

electric-field imaginary part. (C) Computed distribution of the electric field in a horizontal section 25 nm above the Au interface, same wavelength.

In order to get a better physical understanding of the origin of the low wavelength mode presented in **Fig. 3B**, numerical simulations were performed using the Green's tensor method on a single Au NP, deposited either onto glass coated with 50 nm Au film or on glass coated with 50 nm Au and post-coated with graphene (glass/Au/graphene) with illumination in normal incidence to the substrate. The thickness of the graphene layer was chosen as 1 nm since experimentally we determined ≈3 monolayers. The optical constants of Au were taken from Johnson and Christy (Johnson and Christy, 1972). The numerical extinction spectra for the different interfaces with Au NPs of 50 nm in height and varying particle diameter (80 nm, 110 nm and 140 nm) are seen in **Fig. 4** for the systems with and without graphene spacer.

**Fig. 4A** displays a sharp peak at λ=520 nm for both systems independent on the presence of graphene as spacer. The electric field maps (**Fig. 4B** and **4C**) indicate that this mode corresponds to a dipolar localized surface plasmon whose hot spots (zones of high near-field intensities) are "pushed" to the top corners of the NPs, thus at the interface with air. The lower optical index of air compared to that of substrates made of higher index dielectric materials (such as glass) would then explain why this mode exhibits a lower resonance wavelength compared to similar gratings fabricated onto $SiO_2$ substrates (Viste et al., 2010). It has been shown that when NPs lie on a dielectric substrate the hot spots of the LSP mode are localized at the low corners of the particles, that is at the substrate surface (Hutter et al., 2013). We believe that the Au film plays a role similar to a mirror on the mode field distribution and that this effect stems from the interference between the field scattered by the NP gratings and the reflection of this very same scattered field by the Au film. However, the numerical simulations did not provide any optical based explanation about the blue shifted plasmon band. It is currently believed that the blue-shift induced by graphene comes from charge transfer between graphene and the Au NPs, which would modify the LSP frequency through a modified free carrier density and plasma frequency.

Motivated by the sharpness of the LSPR mode at 507 nm observed on glass/Au/graphene/Au NPs and its localization at the interface to the surrounding medium, the possibility to use it for sensing applications was investigated. As predicted by Hohenau and Krenn, the LSPR peak sharpness should lead to highly sensitive sensors (Hohenau and Krenn, 2010). The refractive index sensing of the Au NPs/graphene/Au film interface was investigated by recording the wavelength shift when immersed in water/glycerol mixtures at different proportions giving different refractive indexes (n=1.33 for water to 1.47 for glycerol, inbetween water/glycerol mixtures). **Fig. 5A** shows that the position of $λ_1$ is shifting to higher wavelengths with increasing the refractive index. The change in the position of $λ_1$, $Δλ_1$, shows a linear dependency as a function of the refractive index of the surrounding medium. The sensitivity, defined as the ratio of the change in the position of the plasmon band over the change in the refractive index, $dλ/dn$, is determined from the slope of **Fig. 5B** and increases as the thickness/diameter ratio of the plasmonic interface decreases. Graphene-modified Au films coated with Au NPs of 140 nm in diameter exhibit a sensitivity of 139 nm/RIU, whereas the same systems coated with Au NPs of 110 nm and 80 nm display sensitivities of 66 nm/RIU and 34 nm/RIU, respectively. The observed sensitivity of 139 nm/RIU is comparable to other plasmonic structures with a resonance band between 500-600 nm (Malinsky et al., 2001, Mock et al., 2003, Sherry et al., 2005, Khalavka et al., 2009). The 520 nm mode of the system without any spacer exhibits a somewhat lower sensitivity (124 nm/RIU) and a larger peak width (fwhm = 64 nm). It has become common to compare the sensing characterization of a LSPR mode by its figure-of-merit (FoM) defined by the ratio between the sensitivity and the full width at half maximum of the resonance peak (FoM= $(dλ/dn)$/fwhm) with high values of FoM being an indicator for good sensor performance and good readability. For the interface with Au NPs of 140 nm in diameter with graphene spacer, the fwhm=50 nm and results in FoM=2.8 when fitted by a Lorentzian function as compared to only 2.1 without spacer. Thus, it is evident that the graphene spacer enhances both sensitivity and FoM of Au NPs coupled to Au film systems. Moreover, this FoM is higher than those reported for silver triangles ($λ_{peak}$=564 nm; FoM=1.8) (Malinsky et al., 2001), silver cubes ($λ_{peak}$=510 nm; FoM=1.6) (Sherry et al., 2005), silver spheres ($λ_{peak}$=520 nm; FoM=2.2) (Mock et al., 2003) or gold spheres ($λ_{peak}$=530 nm; FoM=1.5) (Underwood and Mulvaney, 1994). It ranks this interface among the highly sensitive LSPR sensors with plasmon band in the visible at 500 nm (Zalyubovskiy et al., 2012). Indeed, most of the interfaces with high FoM (4-16.5) (Verellen et al., 2011, Lodewijks et al., 2012) take advantage of the fact that higher sensitivities are achieved with plasmon bands in the near-infrared of the spectrum (850-1200 nm).

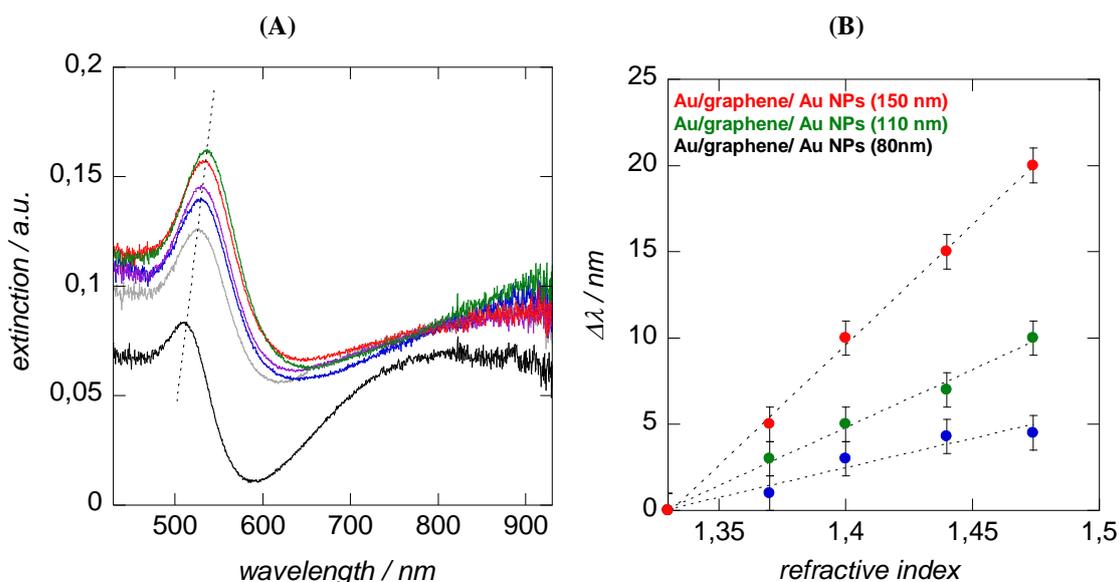

**Fig. 5**: (A) Extinction spectra of the graphene-modified SPR surface decorated with Au NPs array of 140 nm in diameter for different refractive indexes *n* of glycerol/water mixtures: 1.00 (black), 1.33 (grey), 1.37 (blue), 1.40 (magenta), 1.44 (red) 1.47 (green); (B) Shift of the low wavelength LSPR peak depending on the refractive index of the surrounding medium.

In conclusion, in this work the interaction of metallic nanoparticle gratings with gold thin films using graphene as spacer is investigated. Optical extinction measurements allow to evidence that the fabrication of Au NP gratings directly on Au film or separated with trilayered graphene leads to a sharp peak at 520 nm (Au/AuNPs) and 507 nm (Au/graphene/AuNPs), which is almost independent of the size of the nanoparticles. The position of the plasmonic band at 520 nm is in accordance with numerical simulations based on the electromagnetic theory and corresponds to a dipolar LSP mode, which is pushed to the top of the interface and the interface with air. The blue-shift induced by the trilayer graphene spacer could be induced by charge transfer between the graphene layer and the Au NPs gratings. The dipolar LSP mode reveals, however, to be well sensitive to optical index changes of the surrounding medium due to its interface with air. Moreover, the importance in sensing of this LSPR mode is linked to its low fwhm of 50 nm, which results in a FoM of 2.8 at $\lambda_1$ ≈ 507 nm. The role of the graphene spacer in Au NPs coupled to Au film systems is clearly evidenced to both increase the sensitivity and decrease the FWHM of the LSPR peak which leads to a large improvement of the FoM from 2.1 to 2.8, that is to say about 33 %. Besides, this study provides a further understanding of systems based on arrays of resonant metallic NPs coupled to metallic films. From a practical point of view, it opens avenues to engineering in a controlled and predictable way the spectral properties of metallic NPs-based systems to reinforce their applicability especially for sensing applications. This first study paves thus the way for highly sensitive sensors and should lead to further studies in order to optimize both the number of graphene layers and the NP size.

ASSOCIATED CONTENT
**Supporting Information**. Supporting Information contains details about graphene preparation and transfer. Structural characterization of the graphene before and after its transfer was performed using Raman and XPS measurements (see Figure S1). SPR investigations compared to numerical simulations allowed to determine the number of graphene layers (see Figure S2).

**Acknowledgments**
Financial support of NanoMat (www.nanomat.eu) by the "Ministère de l'enseignement supérieur et de la recherche," the "Conseil régional Champagne-Ardenne," the "Fonds Européen de Développement Régional (FEDER) fund," and the "Conseil général de l'Aube" is acknowledged. The EU-ERDF via the Interreg IV Programme (project "Plasmobio) are also gratefully acknowledged for financial support. S. S thanks the Institut Universitaire de France (IUF). T. M thanks the DRRT (Délégation Régionale à la Recherche et à la Technologie) of Champagne-Ardenne, the Labex ACTION project (contract ANR-11-LABX-01-01) and the CNRS via the chaire « optical nanosensors » for financial support.